\documentclass[12pt]{article}
\usepackage[dvips]{graphics}
\usepackage{graphicx}
\title{Time in relativistic and nonrelativistic quantum mechanics}
\author{Hrvoje Nikoli\'c \\
Theoretical Physics Division, Rudjer Bo\v{s}kovi\'{c} Institute, \\
P.O.B. 180, HR-10002 Zagreb, Croatia \\
{\normalsize e-mail: hrvoje@thphys.irb.hr} \\
\makebox[1in]{} \\
}
\date{\today}
\begin{document}
\maketitle

\begin{abstract}
The kinematic time operator can be naturally defined in relativistic
and nonrelativistic quantum mechanics (QM) by treating time on an equal
footing with space. The spacetime-position operator acts in the
Hilbert space of functions of space and time. 
Dynamics, however, makes eigenstates of the time operator unphysical.
This poses a problem for the standard interpretation of QM 
and reinforces the role of alternative interpretations such as the
Bohmian one. The Bohmian interpretation, despite of being nonlocal
in accordance with the Bell theorem, is shown to be relativistic covariant.
\end{abstract}
\vspace*{0.5cm}
\maketitle
%\pacs{03.65.Ta, 03.65.Pm}
%Foundations of quantum mechanics; measurement theory,
%Relativistic wave equations

\section{Introduction}

The problem of time-operator in quantum mechanics (QM) is an old,
still unsolved problem (see, e.g., \cite{busch,bostr,nikmyth}
for reviews). In its most elementary form, the problem can be stated
as an old observation \cite{pauli} that,
if the Hamiltonian $\hat{H}$ is bounded from below,
there can be no self-adjoint time-operator
$\hat{T}$ satisfying the canonical commutation relation
\begin{equation}\label{e1}
[\hat{T},\hat{H}]=-i\hbar .
\end{equation}
This is especially problematic for relativistic QM, because
in a relativistic theory one expects that time should be treated
on an equal footing with space.
Guided by the idea that time should be treated
on an equal footing with space, in this paper we propose a 
simple solution of that problem. Essentially, when acting
on a wave function $\psi({\bf x},t)$, the time operator
acts as a multiplication with the parameter $t$. To make it meaningful,
the Hilbert space in which the physical operators act
must be enlarged from the space of functions $f({\bf x})$ of ${\bf x}$
to the space of functions $f({\bf x},t)$ of both ${\bf x}$ and $t$.
The quantity $|\psi({\bf x},t)|^2 d^3x \, dt$ is then naturally interpreted
as the probability that the particle will be found in the (infinitesimal)
spacetime volume $d^3x \, dt$. 

Some forms of this idea
have also been proposed in the context of
proper-time quantization of relativistic particles 
\cite{stuc}, relativistic quantum measurements
\cite{rov}, and measurement events \cite{bostr}.
However, even though such a construction works perfectly well
on the kinematic level (i.e., without using the wave equation
of motion for $\psi({\bf x},t)$), the common and unsolved problem 
of all these previous attempts is the fact that it does not 
work on the dynamical level, essentially because eigenstates
of the time operator are not solutions of the equation of motion.
In this paper we propose a solution of that problem as well.
Moreover, we discuss how several other foundational problems
of QM are naturally solved within this approach. 
More specifically, the list of advantages of this approach
includes the following:
\begin{itemize}
\item  Space and time operators are defined in a relativistic covariant
way.
\item A relativistic covariant probabilistic interpretation
of solutions of the Klein-Gordon equation is given.
\item The standard interpretation of transition amplitudes
in terms of transition probabilities {\em per unit time}
is now better founded in basic axioms of QM.
\item The fact that not all interpretations of QM are compatible
with the fact that eigenstates of the time operator are unphysical 
provides a new guiding principle towards a 
resolution of notorious interpretational ambiguities
related to the measurement problem in QM.
\item In the Bohmian interpretation the dynamical absence of time-eigenstates
does not represent a fundamental problem. 
\item Even though the Bohmian interpretation
is not local, it is shown to be relativistic covariant. 
\end{itemize}
In the rest of the paper we explain these properties in more detail.
From now on, we use natural units $\hbar=c=1$ and the
relativistic metric signature $(+,-,-,-)$.

\section{Kinematics}

Let us start with a purely kinematic analysis.
We introduce the standard relativistic notation $x=\{ x^{\mu} \}$, 
$\mu=0,1,2,3$, where $x^0\equiv t$ and $\{x^1,x^2,x^3\}\equiv {\bf x}$.
The kinematic 4-momentum operator is
\begin{equation}\label{e2}
\hat{p}_{\nu}=i\partial_{\nu} ,
\end{equation}
which satisfies the covariant canonical commutation relations
\begin{equation}\label{e3}
[x^{\mu},\hat{p}^{\nu}]=-ig^{\mu\nu} .
\end{equation}
The time operator is simply the $x^0$ component of $x^{\mu}$.
Similarly, the space-position operator is the space-component 
${\bf x}$ of $x^{\mu}$. Thus, unlike the Newton-Wigner position operator
\cite{newt}, our position operator is relativistic covariant
by being a space-component of a Lorentz 4-vector.
The time-component of (\ref{e2}) is the kinematic energy operator
\begin{equation}\label{e4}
\hat{E}=i\frac{\partial}{\partial t} ,
\end{equation}
which may serve as an energy operator even in the nonrelativistic case.
Thus, instead of the inconsistent relation (\ref{e1}), Eq.~(\ref{e3})
leads to a consistent one
\begin{equation}\label{e5}
[t,\hat{E}]=-i .
\end{equation}
Unlike the dynamical Hamiltonian operator $\hat{H}$, the kinematic
energy operator $\hat{E}$ is not bounded from below.
The operators (\ref{e2}) act on the space of functions $f(x)$ of $x$.
The natural scalar product on this space is
$\langle\psi|\psi'\rangle = \int d^4x \, \psi^*(x)\psi'(x)$.
In particular, if $\psi$ is normalized such that 
$\langle\psi|\psi\rangle =1$, then the quantity
\begin{equation}\label{e6}
dP=|\psi(x)|^2d^4x
\end{equation}
is naturally interpreted as the probability of finding the particle
in the (infinitesimal) 4-volume $d^4x$. 

At first sight, (\ref{e6}) may seem to be incompatible with the usual
probabilistic interpretation in 3-space 
\begin{equation}\label{e7}
dP_{(3)}=|\psi({\bf x},t)|^2d^3x .
\end{equation}
Nevertheless, (\ref{e6}) is compatible with (\ref{e7}). If (\ref{e6})
is the fundamental {\it a priori} probability, then (\ref{e7})
is naturally interpreted as the conditional probability corresponding to the case
in which one knows that the particle is detected at time $t$.
More precisely, $\psi$ in (\ref{e6}) and (\ref{e7}) have different normalizations,
so a more precise form of (\ref{e7}) is
\begin{equation}\label{e8}
dP_{(3)}=\frac{|\psi({\bf x},t)|^2 d^3x}{N_t},
\end{equation}
where 
\begin{equation}\label{e9}
 N_t=\int d^3x |\psi({\bf x},t)|^2 
\end{equation}
is the normalization factor. If $\psi$ is normalized such that (\ref{e6}) is valid,
then (\ref{e9}) is also the marginal probability that the particle will be found
at $t$. Of course, in practice a measurement allways lasts a finite time $\Delta t$
and the detection time $t$ cannot be determined with perfect accuracy. 
Thus, (\ref{e8}) should be viewed as a limiting case in which
the fundamental probability (\ref{e6}) is averaged over a very small $\Delta t$.
More precisely, if the particle is detected 
between $t-\Delta t/2$ and $t+\Delta t/2$, then  
(\ref{e8}) is the probability of different 3-space positions of the particle detected 
during this small $\Delta t$.

Can the probabilistic interpretation (\ref{e6}) be verified experimentally?
In fact, it already is! In practice one often measures cross sections 
associated with scaterring experiments or decay widths and lifetimes
associated with spontaneous decays of unstable quantum systems. 
These experiments agree with standard theoretical predictions. Our point is
that these standard theoretical predictions actually use (\ref{e6}),
although not explicitly. Let us briefly explain it. (A complete explanation
requires an explicit and careful account of all normalization factors.
Since this is not essential for understanding the basic idea,
a complete analysis will be presented elsewhere.) 
The basic theoretical tool
in these predictions is the transition amplitude $A$. Essentially,
the transition amplitude is the wave function (usually Fourier transformed
to the 3-momentum space) at $t\rightarrow\infty$, calculated by 
assuming that the wave function at $t\rightarrow -\infty$ is known.
Due to energy conservation one obtains $A\propto\delta(E_{\rm in}-E_{\rm fin})$,
where $E_{\rm in}$ and $E_{\rm fin}$ are the initial and final energy, respectively.
Thus, the transition probability is proportional to
$|A|^2\propto[\delta(E_{\rm in}-E_{\rm fin})]^2=(T/2\pi)\delta(E_{\rm in}-E_{\rm fin})$,
where $T=\int dt =2\pi \delta(E=0)$. Since $T$ is infinite, this transition
probability is physically meaningless. The standard interpretation 
(see, e.g., \cite{schiff} for the nonrelativistic case or \cite{halz} for the
relativistic case), which agrees with experiments, 
is that the physical quantity is $|A|^2/T$ and that this quantity is (proportional to)
the transition probability {\em per unit time}. But this is essentially the same
as our equation (\ref{e6}) which says that $\int d^3x |\psi|^2$ is not 
probability itself, but probability {\em per unit time}. Although 
the interpretation of $|A|^2/T$ as probability per unit time 
may seem plausible even without explicitly postulating (\ref{e6}),
without this postulate such an interpretation of $|A|^2/T$
is at best heuristic and cannot be strictly derived from other basic
postulates of QM, including (\ref{e7}).
In this sense, {\em the standard interpretation of transition amplitudes
in terms of transition probabilities per unit time
is better founded in basic axioms of QM if (\ref{e6}) is also adopted
as one of its axioms}.

\section{Dynamics and generalization to many particles}

So far we have been explicitly discussing only the kinematics. Now let us include
the dynamics.  For simplicity, we shall only consider particles without spin.
Thus, the wave function of a free relativistic particle satisfies the Klein-Gordon
equation
\begin{equation}\label{e10}
 [\hat{p}^{\mu}\hat{p}_{\mu}-m^2]\psi(x)=0 .
\end{equation}
The interaction with a background field $A^{\mu}(x)$ can also be included through the
substitution $\hat{p}^{\mu}\rightarrow \hat{p}^{\mu}+eA^{\mu}(x)$,
but the interaction will not influence our conclusions, so we consider only the free case.
By writing $\psi(x)=e^{-imt}\psi_{\rm NR}({\bf x},t)$ and taking the nonrelativistic
limit of (\ref{e10}), one finds that $\psi_{\rm NR}$ satisfies the nonrelativistic
Schr\"odinger equation. Hoping that it will not cause confusion, in the rest of the
discussion we omit the label ``NR'' in $\psi_{\rm NR}$,
so the nonrelativistic Schr\"odinger equation can be written as
\begin{equation}\label{e11}
\hat{H}\psi({\bf x},t)=\hat{E}\psi({\bf x},t) ,
\end{equation}
where $\hat{H}=\hat{\bf p}^2/2m$ is the dynamical Hamiltonian operator that 
acts in the space of functions of ${\bf x}$, while $\hat{E}$
given by (\ref{e4}) is the kinematic energy operator that acts 
in the space of functions of $t$.
Since $\hat{H}$ is positive definite, Eq.~(\ref{e11}) implies that wave functions
of the form $e^{-iEt}\chi({\bf x})$ may be solutions only if $E\geq 0$.
This means that the $\delta$-function 
\begin{equation}\label{e12}
\delta(t)=\int_{-\infty}^{\infty}\frac{dE}{2\pi}e^{-iEt} 
\end{equation}
cannot be constructed from solutions of (\ref{e11}). In other words,
in the nonrelativistic case eigenstates of the time operator are {\em not physical}.

Can this problem be avoided by using relativistic QM? Eq.~(\ref{e10})
contains solutions with both positive and negative energies $E$.
Nevertheless, the spectrum of allowed energies is restricted again,
by the condition $|E|\geq m$. Moreover, even the case $m=0$ is 
problematic, because it is usually assumed that (for uncharged particles)
only positive energy
solutions of (\ref{e10}) are physical. (There are many ways to explain 
why only positive $E$'s are physical. 
One way is to observe that only with that restriction 
Eq.~(\ref{e10}) can be reduced to Eq.~(\ref{e11}) in the nonrelativistic limit.
Another way is to use relativistic quantum field theory, where physical states are obtained
by acting with creation operators $\hat{a}^{\dagger}({\bf p})$ 
on the vacuum, which create states with positive $E$ only \cite{ryder}.)
 
Thus, we see that eigenstates of the time operator cannot be constructed
from physical solutions of the dynamical equations of motion. Does it mean
that our kinematic time operator is physically inconsistent?
For example, the time operator cannot be represented in terms of physical
states as $\hat{t}=\int dt |t\rangle t \langle t|$. Nevertheless, this fact by itself
does not yet represent a physical problem, as long as this operator still exists
kinematically. Still, a physical problem would occur if this would imply that
time cannot be {\em measured}. But this leads us to a much 
more controversial aspect of QM -- the theory of measurement. 
 
Before discussing the problems related to quantum measurements, let us 
first clarify some additional, less problematic, aspects of our approach. Another problem with 
physical solutions of (\ref{e10}) and (\ref{e11}) is that they cannot be localized in time,
i.e., they satisfy $\int_{-\infty}^{\infty} dt |\psi|^2 = \infty$. 
Strictly speaking, this means that $\psi$ cannot be normalized such that 
$\langle\psi|\psi\rangle =1$. Nevertheless, we do not see this as a real physical problem.
After all, no experiment lasts an infinite time. And even if it does, one can
allways introduce a large but finite time-cutoff $T$ such that
$\int_{-T/2}^{T/2} dt |\psi|^2$ is finite, and put $T\rightarrow\infty$ 
at the end of calculation. Indeed, as we already discussed above, 
a regularization of that kind is a standard procedure in calculations of 
transition probabilities per unit time and no serious problems have been
found so far.

Another aspect that we want to discuss briefly is the generalization to
wave functions of $n$ particles. 
As we already pointed out, the main idea is to treat time
on an equal footing with space. Thus, each particle has its own space 
position ${\bf x}_a$, $a=1,\ldots,n$, as well as its own time coordinate
$t_a$. Therefore, the wave function is of the form $\psi(x_1,\ldots,x_n)$,
which is a many-time wave function \cite{tomon}. Then (\ref{e6}) generalizes to
\begin{equation}\label{e6n}
dP=|\psi(x_1,\ldots,x_n)|^2 d^4x_1 \cdots d^4x_n .
\end{equation}
Hence, if the first particle is detected at $t_1$, second particle at $t_2$, etc., 
then Eq.~(\ref{e8}) generalizes to
\begin{equation}\label{e7n}
dP_{(3n)}=\frac{|\psi({\bf x}_1,t_1,\ldots,{\bf x}_n,t_n)|^2 d^3x_1 \cdots d^3x_n}
{N_{t_1,\ldots,t_n}} ,
\end{equation}
where 
\begin{equation}\label{e8n}
N_{t_1,\ldots,t_n}=\int |\psi({\bf x}_1,t_1,\ldots,{\bf x}_n,t_n)|^2 d^3x_1 \cdots d^3x_n .
\end{equation}
If these particles do not interact, then (in the relativistic case) $\psi$ satisfies
$n$ Klein-Gordon equations $ [\hat{p}_a^{\mu}\hat{p}_{a\mu}-m^2_a]\psi=0$,
one for each $a$. They can also be summed to give a single $n$-particle
Klein-Gordon equation
\begin{equation}\label{e10n}
\sum_a [\hat{p}_a^{\mu}\hat{p}_{a\mu}-m^2_a] \psi(x_1,\ldots,x_n) =0.
\end{equation}
By taking the nonrelativistic limit and considering the coincident case
$t_1=\cdots=t_n\equiv t$, one finds that these equations reduce to the standard
single-time nonrelativistic equations of $n$-particle systems.

\section{Quantum measurements and the
Bohmian interpretation}

Now let us discuss our final and the most difficult issue -- the issue of quantum
measurements. According to the standard interpretation of QM, 
when a physical observable is measured, then the wave function collapses
to an eigenstate of the operator describing this observable.
However, as we have seen, eigenstates of the time operator are unphysical.
At first sight it may seem to be a serious drawback of our approach.
Nevertheless, we argue that it is actually a {\em virtue}, rather than a drawback.
Namely, even though such a time operator is not consistent within the
standard interpretation of QM, it may still be consistent within some of
the alternative interpretations. Usually, the question of ``correct''
interpretation is viewed as something that belongs to philosophy, rather than 
science. But now we propose a new scientific criterion for distinguishing
between acceptable and unacceptable interpretations; {\em in an acceptable 
interpretation the fact that time eigenstates are not physical should not
imply that time itself is unmeasurable}.

It is beyond the scope of this paper to discuss all acceptable interpretations.
Instead, we concentrate on one such interpretation -- the Bohmian one
\cite{bohm}. According to this interpretation, in experiments
we actually do not detect wave functions,
but pointlike particles that move deterministically through spacetime and exist
as objects separated from (although guided by) the wave function.
This explains why the spacetime position of a particle makes physical
sense even without eigenstates of the time operator (see also
\cite{leav}). But how can time be measured? 
According to the Bohmian interpretation, 
all quantum measurements eventually reduce to measurements
of 3-space positions of particles of the measuring apparatuses \cite{bohm},
which then also applies to the measurement of ``time'' by a real clock.
(For explicit models of a clock see, e.g., \cite{wigner}.)

The Bohmian interpretation has also other advantages (see, e.g., 
\cite{bohmPRL}). 
Moreover, many objections against this interpretation have been found
to be unjustified. In particular, contrary to frequent claims, it was 
found that the Bohmian interpretation {\em does} have a practical use
\cite{lopr}, that the Bohmian particle velocities {\em are} measurable
\cite{bohmweak} (in the sense of weak measurements \cite{ahar}),
and that (an improved version of) Bohmian mechanics
{\em can} describe particle creation and destruction,
by using either quantum field theory \cite{QFTcr} or string theory 
\cite{nikstr1,nikstr2}. In addition, it is sometimes objected
that the Bohmian interpretation is nonlocal, but this is not really a
valid argument against this particular interpretation because {\em any} theory 
(compatible with QM) that assumes that single reality exists even without 
measurements must necessarily be nonlocal \cite{bell}.
 
Finally, due to nonlocality, it is frequently objected that this interpretation
is not consistent with relativity. Nevertheless, various partial steps towards 
a relativistic-covariant formulation of the Bohmian interpretation
of many-particle systems have been done in \cite{relatbohm} and \cite{nikstr1}.
Here we make a synthesis of these partial steps 
with the results of the present paper
to show in a simple and concise way
that the Bohmian interpretation is indeed relativistic covariant,
despite of being nonlocal. 

Each Bohmian particle has a trajectory $X^{\mu}_a(s)$, where $s$ is an auxiliary
scalar parameter along the trajectories. Since $\psi(x_1,\cdots,x_n)$
does not depend on $s$, by writing $\psi=|\psi|e^{iS}$
one finds that (\ref{e10n}) implies a relativistic conservation equation
\begin{equation}\label{b1}
 \frac{\partial |\psi|^2}{\partial s} + \sum_a \partial_{a\mu}(|\psi|^2 v^{\mu}_a) =0 ,
\end{equation}
where 
\begin{equation}\label{b2}
v^{\mu}_a(x_1,\cdots,x_n) = -\partial_a^{\mu}S(x_1,\cdots,x_n) .
\end{equation}
Therefore, it is consistent to postulate that the trajectories satisfy
\begin{equation}\label{b3}
 \frac{dX^{\mu}_a(s)}{ds}=v^{\mu}_a(X_1(s),\cdots,X_n(s)) .
\end{equation}
Namely, if an ensemble of particles has the distribution (\ref{e6n})
for some initial $s$, then (\ref{b1}) and (\ref{b3}) guarantee
that it will have the same distribution for any $s$.
The Bohmian equation of motion (\ref{b3}) is nonlocal
because it says that the velocity of a particle for some value of $s$ depends 
on the positions of all other particles for the same value of $s$.
Nevertheless, it is clear that
this equation is manifestly relativistic covariant. In fact, the auxiliary parameter
$s$ can be completely eliminated and (\ref{b3}) can be rewritten 
as an equation that determines particle trajectories in spacetime
without any referring to the parameter $s$ \cite{relatbohm}.
Finally, we also note that the velocities (\ref{b3}) can be even measured in the sense of
weak measurements, completely analogously to the weak measurements
of nonrelativistic Bohmian velocities \cite{bohmweak}.
To conclude, all this shows that {\em the Bohmian interpretation is well
motivated, relativistic covariant, and compatible with the relativistic
invariant probabilistic interpretation (\ref{e6n})}.

\section*{Acknowledgements}

This work was supported by the Ministry of Science of the
Republic of Croatia under Contract No.~098-0982930-2864.

\end{document}